\documentclass[12pt,preprint,natbib]{aastex}

\input epsf

\def\s2n{S^{\prime}/N}

\def\jco{{J=1-0 $^{13}$CO}}
\def\co{$^{13}$CO}

\begin{document}
\title{Structure Function Scaling in the Taurus and Perseus Molecular Cloud Complexes.}

\author{Paolo Padoan}
\affil{Jet Propulsion Laboratory, California Institute of Technology, 
4800 Oak Grove Drive, MS 169-506, Pasadena, CA 91109}
\email{padoan@jpl.nasa.gov}

\author{Stanislav Boldyrev}
\affil{Institute for Theoretical Physics, Santa Barbara, CA 93106}

\author{William Langer}
\affil{Jet Propulsion Laboratory, California Institute of Technology, 
4800 Oak Grove Drive, MS 169-506, Pasadena, CA 91109}

\and 

\author{{\AA}ke Nordlund}
\affil{Astronomical  Observatory and Theoretical  Astrophysics  Center,  
Juliane  Maries  Vej  30,  DK-2100 Copenhagen, Denmark, aake@astro.ku.dk}

\begin{abstract}

We compute the structure function scaling of the integrated intensity
images of two \jco\ maps of Taurus and Perseus. The scaling exponents
of the structure functions follow the velocity
scaling of super--sonic turbulence, suggesting that turbulence plays
an important role in the fragmentation of cold interstellar clouds.
The data also allows to verify the validity of the two basic assumptions
of the hierarchical symmetry model, originally proposed for the derivation
of the velocity structure function scaling. This
shows that the same hierarchical symmetry holds for the projected
density field of cold interstellar clouds.

\end{abstract}

\keywords{
turbulence -- ISM: kinematics and dynamics -- radio astronomy: interstellar: lines
}

\section{Introduction}

Due to the complexity of the Navier--Stokes equations, mathematical work
on turbulence is often inspired by experimental and observational
measurements. Since geophysical and laboratory flows are predominantly
incompressible, turbulence studies have been limited almost entirely
to incompressible flows (or to infinitely compressible ones, described
by the Burgers equation). Little attention has been paid to
highly--compressible, or super--sonic turbulence.

The cold interstellar medium (ISM) of galaxies such as the Milky Way
is both highly turbulent and highly super--sonic, within
a range of scales from hundreds of parsecs to approximately a
tenth of a parsec. Large observational surveys of the cold ISM of our
galaxy and of dust within it have become available in the last few years
and new surveys of unprecedented sensitivity and resolution will
be obtained in the near future by FIR satellites such as SIRTF
and Herschel.

The cold ISM provides a good laboratory for super--sonic
turbulence and results from new observational surveys should motivate
new mathematical work.
Furthermore, the interpretation of the astronomical data requires
a knowledge of basic properties of super--sonic turbulence.

Previous works have tried to investigate the properties of ISM
turbulence by estimating the second order structure function
or the power spectrum of the velocity field in dark clouds
\citep{Kleiner+Dickman87,Hobson92,Miesch+Bally94,%
Brunt+Heyer2002calibration,Brunt+Heyer2002results},
as sampled by molecular emission lines,
with a number of different methods. The centroid velocity at each
map position is used as an estimate of the local radial velocity.
This centroid velocity results from a convolution of density,
velocity and excitation temperature along the line of sight
and is not easily related to the three dimensional velocity structure.
Other works have instead studied the column density distribution
of dark clouds, estimated from the integrated intensity of
molecular emission lines or from dust thermal emission. Fractal
dimensions have been estimated
\citep{Beech87,Bazell+Desert88,Scalo90,Dickman+90,%
Falgarone+91,Zimmermann+92,Henriksen91,Hetem+Lepine93,Vogelaar+Wakker94,%
Elmegreen+Falgarone96}.
Multifractal \citep{Chappell+Scalo01}
and wavelet \citep{Langer+93} analysis have also been
proposed as a way of characterizing the projected density structure
of molecular clouds.

The multifractal analysis applied by \cite{Chappell+Scalo01} to
dust continuum images is related to the scaling of the moments of
the projected density. In the present work we study the
scaling of integrated intensity differences (structure functions)
of the J=1--0 transition of $^{13}$CO from the Taurus and the
Perseus molecular cloud complexes. If we had computed the moments
of intensity, instead of intensity differences, then or analysis
would be equivalent to the multifractal analysis by \cite{Chappell+Scalo01}.

We have already shown in previous works that super--sonic turbulence
in a roughly isothermal gas, such as the cold ISM, generates a complex
density field with density contrasts of several orders of magnitude
and with statistical properties consistent with observational
data from molecular clouds
\citep{Padoan+98cat,Padoan+99per,Padoan+Nordlund99mhd}.
%(Padoan et al. 1998, 1999, 2001; Padoan \&
%Nordlund 1999).
We usually refer to this process
as {\it turbulent fragmentation}. Turbulent fragmentation of star forming
clouds is unavoidable, since super--sonic turbulent motions are
ubiquitously observed. The results of this work provide further
evidence of the importance of turbulent fragmentation in star
forming clouds.

In \S~2 we compute the relative scaling of the structure functions
and in \S~3 we discuss our results. The hierarchical
structure model is briefly presented in \S~4 and its validity
for our data is verified with the so called $\beta$ and $\gamma$
tests. We draw our conclusions in \S~5.

\section{Structure Functions of Projected Density Field}

We use two observational \jco\ spectral maps of the Perseus MC complex
\citep{Billawala+97} and of the Taurus MC complex \citep{Mizuno+95}.
These MC complexes have an extension of approximately 30~pc, and rms radial
velocity $\sigma_v=2.0$~km/s in Perseus and $\sigma_v=1.0$~km/s in
Taurus. The rms sonic Mach number is therefore approximately 14 in Perseus
and 7 in Taurus, assuming a gas kinetic temperature of approximately
10~K. The angular resolution is approximately 0.09~pc in the Perseus map,
assuming a distance of 300~pc and 0.08~pc in the Taurus map,
assuming a distance of 140~pc.

In each map we compute an image of the integrated intensity, $I({\bf r})$,
defined as:
\begin{equation}
I({\bf r})={\sum_v T(v,{\bf r})\,dv},
\end{equation}
where $T(v,{\bf r})$ is the antenna temperature at the velocity channel
$v$ and map position ${\bf r}$. The integrated intensity is set to
zero in each map position with peak antenna temperature smaller than five
times the rms noise averaged over the whole map. The integrated intensity
of \jco\ is roughly proportional to the total gas column density since
in most lines of sight of the two maps analyzed here this transition is
not optically thick. However, the conversion from \jco\ integrated
intensity and gas column density also depends on the abundance ratio
between \co\ and H$_2$, and on the precise spatial distribution of
the excitation temperature (see \cite{Padoan+98co} for a detailed
computation of the effect of the spatial distribution of excitation
temperature on the estimation of the \jco\ column density).

The structure functions of the integrated intensity image $I({\bf r})$
are defined as:
\begin{equation}
S_p(r)=\langle |I({\bf r'})-I({\bf r'}+{\bf \Delta r'})|^p\rangle=
          \langle |\Delta I|^p\rangle
\label{mom}
\end{equation}
where $p$ is the order and the average is extended to all map positions
${\bf r'}$ and all position differences ${\bf \Delta r'}$ such that
$|{\bf \Delta r'}|=r$.

It is possible to compute the moments by first deriving the probability
density function (PDF) of $\Delta I$, $P(\Delta I,r)$, from the observational
map, and then computing:
\begin{equation}
S_p(r)={\int_{\Delta I} \Delta I^p P(\Delta I,r) d(\Delta I)}
\label{pdf}
\end{equation}
The PDF can be obtained by smoothing appropriately the histogram
obtained from the observational data, as suggested by \cite{Leveque+She97}.
Here, however, we prefer to average the moments directly over
the whole observational sample, as in (\ref{mom}), without first
deriving the PDF.

The structure functions are plotted in Figure~1, up to the order
$p=20$. They are well approximated by a power law, at least at low
orders, between approximately 0.3 and 3 pc, for both Taurus and Perseus.
We have performed a least square fit in that range of scales for
all orders. The least square fits are shown in Figure~1 as solid lines.
We call $\eta(p)$ the exponent of the power law fit of the $p$
order moment\footnote{The Greek letters $\zeta$ and $\tau$ are usually
reserved for the exponents of the moments of velocity differences and
energy dissipation respectively.}:
\begin{equation}
S_p(r)\propto r^{\eta(p)}
\label{}
\end{equation}
We then compute the scaling exponents relative
to the third order, $\eta(p)/\eta(3)$, following the idea of
extended self--similarity \citep{Benzi+93,Dubrulle94}.
%(Benzi et al. 1993; Dubrulle 1994).

Values of $\eta(p)/\eta(3)$ for both Taurus and Perseus are plotted
in Figure~2, up to the 20th order, and are compared with those predicted
for the scaling of the structure functions of the velocity field
($\zeta(p)/\zeta(3)$) by the Kolmogorov model \citep{Kolmogorov41},
the She \& L\'{e}v\^{e}que model (formula (\ref{s-l}) below) and
the model by \cite{Boldyrev2002}. The latter is an extension of
the She--L\'{e}v\^{e}que model to super--sonic turbulence, verified with
numerical simulations of super--sonic turbulence by
\citep{Boldyrev+2002structure,Boldyrev+2002scaling}.
The scaling exponents of the structure functions of the integrated
intensity images of Taurus and Perseus follow very closely the
velocity scaling found in \cite{Boldyrev2002}, within 5\% accuracy up to
the 20th order for Taurus and within 10\% accuracy for
Perseus.

For practical purposes it may be useful to know the intrinsic value of
the scaling exponents, not just that relative to the third order one.
For Taurus we find $\eta(3)=1.10$ and for Perseus $\eta(3)=1.18$, both
close to the Kolmogorov's value for the velocity structure function,
$\zeta(3)=1.0$. The second order exponents are $\eta(2)=0.77$ for
Taurus and $\eta(2)=0.83$ for Perseus, not far from the value
of $\eta(2)=0.7$  predicted analytically \citep{Boldyrev+2002structure}
from the velocity scaling in \cite{Boldyrev2002}.

\section{Statistical Significance of High Order Moments}

The velocity scalings of laboratory, geophysical or numerical flows are
hardly verified up to the 10th order \citep{Muller+Biskamp2000,%
She+2001}. In the present work we have
used astronomical data to constrain moments beyond the 10th
order. A way to verify the statistical significance
of high order moments is to plot the integrand of the expression
(\ref{pdf}), and verify that at its peak position the PDF is defined
by a significant number of samples. The peak position of the integrand of
(\ref{pdf}) grows with increasing order $p$ and converges to the
value of the cutoff of the PDF. For a finite sample size
the cutoff of the PDF may depend on the sample size, in which case
high order moments would be inaccurate. However, as pointed out in
\cite{Leveque+She97}, if a definite cutoff is present, for example if
it is well defined by a significant number of samples, then
the moment calculation up to any order should be accurate, and the
fast convergence of high order moments is not an artifact of the finite
sample size.

We have plotted in Figure~3, left panel, the tail of the PDF of
$|\Delta I|=|I({\bf r'})-I({\bf r'}+{\bf \Delta r'})|$ from the
Taurus data, for $|{\bf \Delta r'}|=2.6$~pc. Vertical dotted segments
mark the peak position of the integrand of (\ref{pdf}) for
moments of increasing order, from $p=3$ to $p=15$. The
moments converge at $p=15$, and are therefore accurate up to
that order. However, the PDF shows a very sharp cutoff. The
cutoff is definitely significant since it is defined by a few
hundred samples. The `real' PDF describing the physical
process may therefore have a similarly sharp cutoff, in which case
the convergence around the 15th order found in the observational
data would be significant, and orders above the 15th
would be accurately estimated by the present data set.
The moment convergence corresponds to the asymptote of the
function $\eta(p)/\eta(3)$ plotted in Figure~2.
A similar result regarding the PDF cutoff is obtained for
the Perseus data (right panel of Figure~3). However, in this
case the peak of the integrand of (\ref{pdf}) for every order
is very shallow. Although formally the peak position has converged
already at the 5th order (right panel of Figure~3), the moments
converge around the 8th order. The convergence may again be well defined,
as for the Taurus data, since the PDF cutoff is defined by a large sample
size.

%In order to illustrate the moment convergence, we show in Figure~4
%the relative fluctuation of the moments as a function of the sample
%size. The fluctuations are relative to the asymptotic value of the
%largest sample size. The sample size is varied by varying the number
%of directions, ${\bf \Delta r'}$, in the computation of the structure
%functions, among all the available directions.
%$|{\bf \Delta r'}|=0.65$~pc$=8$~pixels has been used in Figure~4.
%A total number of approximately $2\pi\times 8=50$ directions is
%available, which means that in principle we could have reduced
%the sample size by a factor of 50 with this method. We have chosen
%to reduce the sample size by factors of two at each step, down
%to a maximum factor of 16.
%As can be seen in Figure~4, the moments are well converged
%already at a sample size twice smaller than the full sample.
%The convergence of the scaling exponents $\eta(p)$ is even
%better, since they represent the rate of change of the logarithm
%of moments.

\section{The Hierarchical Structure Model for Super--Sonic Turbulence}

The scaling of the velocity structure functions in incompressible
turbulence are best described by the She--L\'{e}v\^{e}que (1994) formula:
\begin{equation}
\frac{\zeta(p)}{\zeta(3)}=\gamma p + C(1-\beta^p)
\label{dubrulle}
\end{equation}
where
\begin{equation}
C=\frac{1-3\gamma}{1-\beta^3}
\label{cod}
\end{equation}
is interpreted as the Hausdorff codimension of the support of
the most singular dissipative structures. In incompressible
turbulence the most dissipative structures are organized in filaments
along coherent vortex tubes with Hausdorff dimension $D=1$ and so
$C=2$. Furthermore, $\beta^3=2/3$ \citep{She+Leveque94}, which
yields the She--L\'{e}v\^{e}que's velocity scaling for incompressible turbulence:
\begin{equation}
\frac{\zeta(p)}{\zeta(3)}=p/9 + 2\left[ 1-\left( \frac{2}{3}\right) ^{p/3}\right]
\label{s-l}
\end{equation}

\cite{Boldyrev2002} has proposed to apply the scaling of (\ref{dubrulle})
to super--sonic turbulence, with the assumption that the
Hausdorff dimension of the support of the most singular dissipative
structures is $D=2$ ($C=1$), since dissipation of super--sonic turbulence
occurs mainly in sheet--like shocks. Using the physical interpretation
of (\ref{dubrulle}) by \cite{Dubrulle94}, $\gamma=1/9$ and
$C=(2/3)/(1-\beta^3)$.  With $C=1$, therefore, $\beta^3=1/3$ and one
obtains the Boldyrev's velocity scaling:
\begin{equation}
\frac{\zeta(p)}{\zeta(3)}=p/9 + 1-\left( \frac{1}{3}\right) ^{p/3}
\label{boldyrev}
\end{equation}

This velocity scaling has been found to provide a very accurate
prediction for numerical simulations of super--sonic and
super--Alfv\'{e}nic turbulence \citep{Boldyrev+2002structure,Boldyrev+2002scaling}.
The computation of the corresponding structure functions for the
gas density distribution is discussed in \cite{Boldyrev+2002structure},
where an exponent $\eta(2)=0.7$ is predicted for
the second order structure function of the projected density.
In this paper we have found the `provoking' result
that the projected density field of super--sonic turbulence in
the ISM has a structure function that is almost indistinguishable
from Boldyrev's velocity scaling.

This result should inspire the mathematical work.
As shown by \cite{Dubrulle94} and by \cite{She+Waymire95}, a
hierarchy of structures producing the scaling relation
(\ref{dubrulle}) can be obtained by a random multiplicative process
with Log--Poisson statistics. The result of this paper suggests
that the density field of super--sonic turbulence is fragmented
by a multiplicative process with Log--Poisson statistics.

\subsection{The $\beta$ and $\gamma$ Tests for the Hierarchical Model}

In the previous section the scaling of the structure
functions of integrated intensity of two maps of the Taurus and Perseus
molecular cloud complexes has been compared with the velocity
scaling of super--sonic turbulence. A different approach would be
to use observational, experimental or numerical data to determine
the parameters $\beta$ and $\gamma$ from a fit to (\ref{dubrulle}).
There is, however, an independent method to first estimate the parameter
$\beta$ and then obtain $\gamma$ from (\ref{dubrulle}),
which also tests the validity of two major
assumptions of the hierarchical structure model at the same time.
This method is presented in \cite{She+2001}; we apply it in the
following.

The basic assumption in the derivation of (\ref{dubrulle}) by
\cite{She+Leveque94} is the existence of the universal
scaling behavior:
\begin{equation}
F_{p+1}(r)=A_p F_p(r)^\beta F_{(\infty)}(r)^{1-\beta}
\label{scaling}
\end{equation}
where
\begin{equation}
F_p(r)=S_{p+1}(r)/S_p(r)=\langle |\Delta I|^{p+1}\rangle/\langle |\Delta I|^p\rangle
\end{equation}
is usually referred to as the $p$th order intensity of fluctuations
and $A_p$ are constants independent of $r$ and found to be
also independent of $p$ in a number of cases.

The $\beta$ test verifies the validity of this basic assumption
with a log-log plot of $F_{p+1}/F_2$ versus $F_p/F_1$. If the
plot is a straight line the assumed hierarchical symmetry
(\ref{scaling}) is satisfied and the data pass the $\beta$
test. The plot is shown in Figure~5, top panel, for both Perseus
(diamond symbols) and Taurus (asterisk symbols). We have combined
together values of $F_p(r)$ for all values of $r$ used to compute
the moment scaling exponents $\eta(p)$ (approximately the range
of scales between 0.3 and 3~pc). Both Taurus and Perseus data
apparently pass the $\beta$ test.

The constants $A_p$ are independent of $p$, as for the hierarchy
of intensity of fluctuations of velocity in the shell model analyzed by
\cite{Leveque+She97}, in the turbulent Couette--Taylor flow
studied by \cite{She+2001} and in laboratory data obtained by
\cite{Chavarria+95}.
The slope of the intensity of fluctuation function plotted in
Figure~5 provides an estimate of the value of $\beta=0.79$ for
Taurus and $\beta=0.78$ for Perseus. This is to be compared with the
value of $\beta=0.69$ of the BSL's velocity scaling for
super--sonic turbulence.

The second basic assumption by \cite{She+Leveque94} is:
\begin{equation}
F_{\infty}\sim S_3^{\gamma}
\label{gamma}
\end{equation}
which allows them to derive (\ref{dubrulle}). The $\gamma$ test
verifies directly the validity of (\ref{dubrulle}) by first
assuming the value of $\beta$ obtained from the $\beta$ test
and then by plotting $\eta(p)-\chi(p,\beta)$ versus
$p-3\chi(p,\beta)$, where $\chi(p,\beta)=(1-\beta^p)/(1-\beta^3)$.
If the plot is a straight line the data pass the $\gamma$ test
and the slope of the plot provides an estimate of the value
of $\gamma$. The plot is shown in Figure~5, bottom panel.
We are able to fit the plot to a straight line
for orders $p>11$ for both Taurus and Perseus. Since $\gamma$
is a property of the very high order moments (see definition
(\ref{gamma})), and since we have obtained a straight line for large moments,
we can say that the Taurus and Perseus data pass the $\gamma$
test. We estimate $\gamma=0.11$ for Taurus and $\gamma=0.06$
for Perseus.

\section{Conclusions}

We have computed the structure functions of the integrated intensity
images of two \jco\ maps of Taurus and Perseus. The structure functions
scale as power laws within the range of scales 0.3--3~pc. The
scaling exponents have been computed up to the 20th order and are
statistically significant at least up to the 15th order in Taurus
and 8th order in Perseus. They are found to follow very closely the
velocity scaling of super--sonic turbulence proposed by \cite{Boldyrev2002},
within 5\% and 10\% accuracy for Taurus and Perseus respectively.

We have verified that the projected density field
(or integrated intensity) of the Taurus and Perseus molecular cloud
complexes can be described by a hierarchical model as the one
proposed by \cite{She+Leveque94} for the velocity structure functions of
incompressible turbulence. We have done so by testing the validity
of the two basic assumptions of the hierarchical model for our
data. The validity of the assumptions of the hierarchical model
means that the integrated intensity images we have analyzed
(an approximate estimate of the projected density of the Taurus and
Perseus molecular cloud complexes) are the result of a multiplicative
process with Log--Poisson statistics \citep{Dubrulle94}.

The complete derivation of the relation between the structure functions of
velocity and projected density is the subject of future work.
However, the close similarity of the structure functions of
projected density in Taurus and Perseus with that of the velocity
field of turbulence provides additional evidence that super--sonic
turbulence is the major factor controlling the density field in the
range of densities and scales sampled by the maps we have analyzed.

It is well established that super--sonic turbulence plays an important
role in the dynamics of the cold ISM
\citep{Larson81,Padoan+98cat,Padoan+99per,Padoan+2001cores,Padoan+Nordlund99mhd}
The statistical properties of this ISM turbulence need to be discovered and
understood in order to elaborate a statistical theory of star formation
\citep{Padoan+Nordlund2002imf}
We have shown in the present paper that this can be achieved with existing
observational data, by studying the projected density field of molecular
clouds. This type of study will be greatly improved
by far infrared imaging of the dust thermal emission from
turbulent ISM clouds, obtained by future satellite missions
such as SIRTF.

\acknowledgements

The work of PP was performed while PP held a National Research
Council Associateship Award at the Jet Propulsion Laboratory,
California Institute of Technology.

\clearpage

%\bibliographystyle{apj}
%\bibliography{apj-jour,padoan,MC,dust}

\clearpage

\onecolumn

{\bf Figure captions:} \\

\begin{figure}
\centerline{
\epsfxsize=9cm \epsfbox{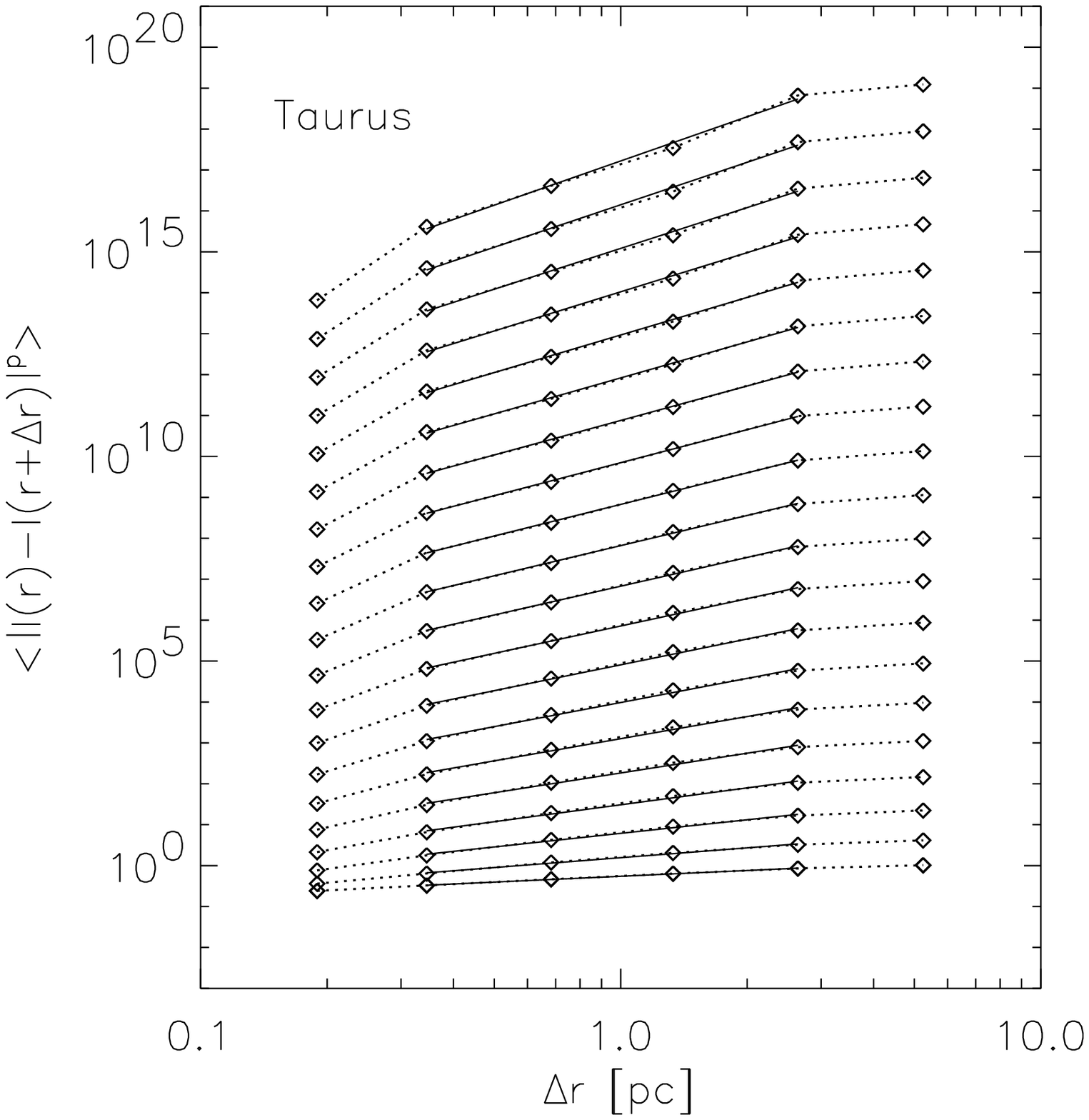}
\epsfxsize=9cm \epsfbox{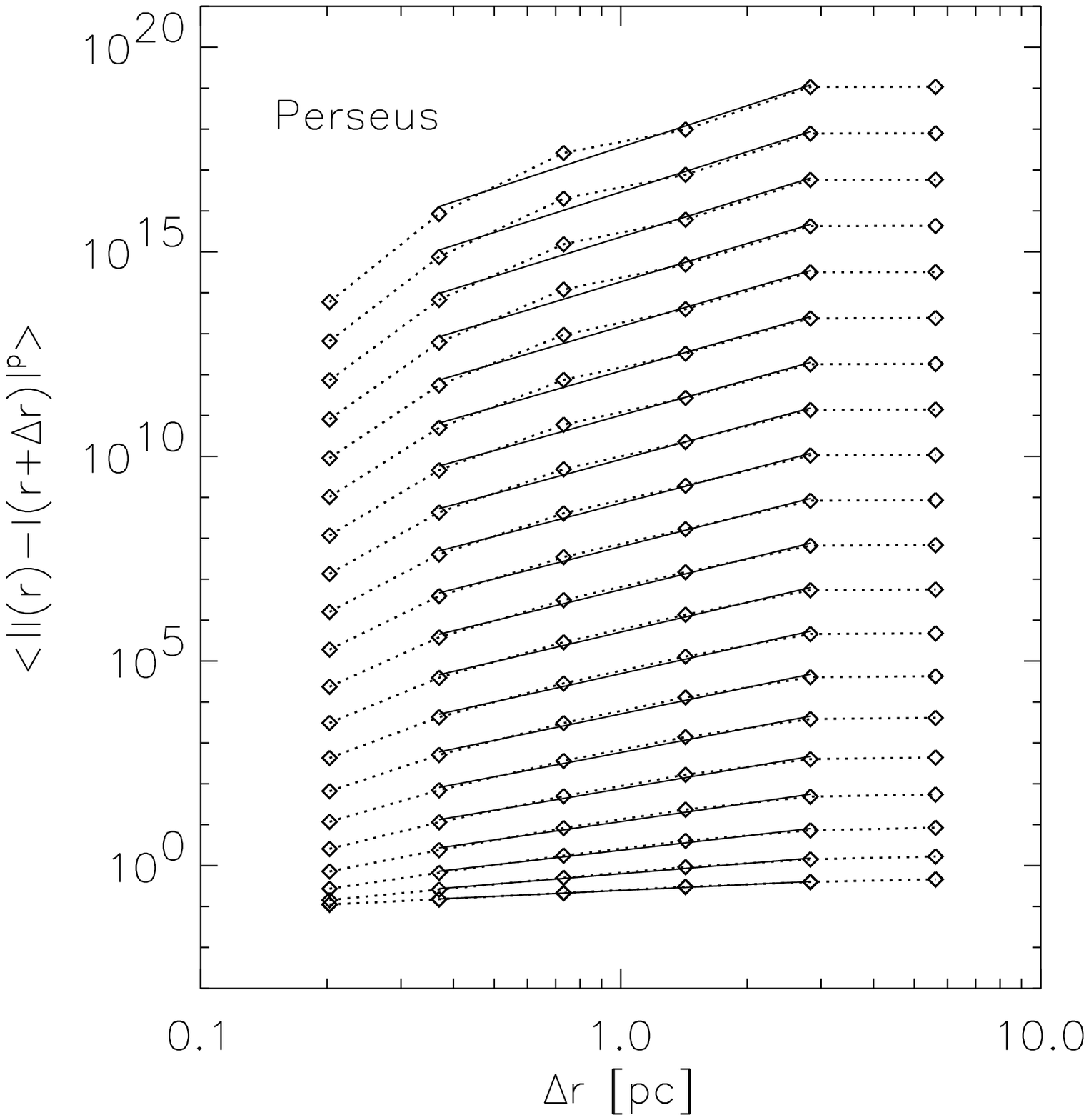}
}
\caption[]{}
\label{fig1}
\end{figure}

{\bf Figure \ref{fig1}:} Structure functions of orders $p=1$ to
to $p=20$ of the integrated intensity \jco\ maps of Taurus
\citep{Mizuno+95} and Perseus \citep{Billawala+97}.
The solid lines are least square fits to the structure
functions in the approximate range 0.3--3~pc, where the structure
functions are well described as power laws. \\

\begin{figure}
\centerline{\epsfxsize=13cm \epsfbox{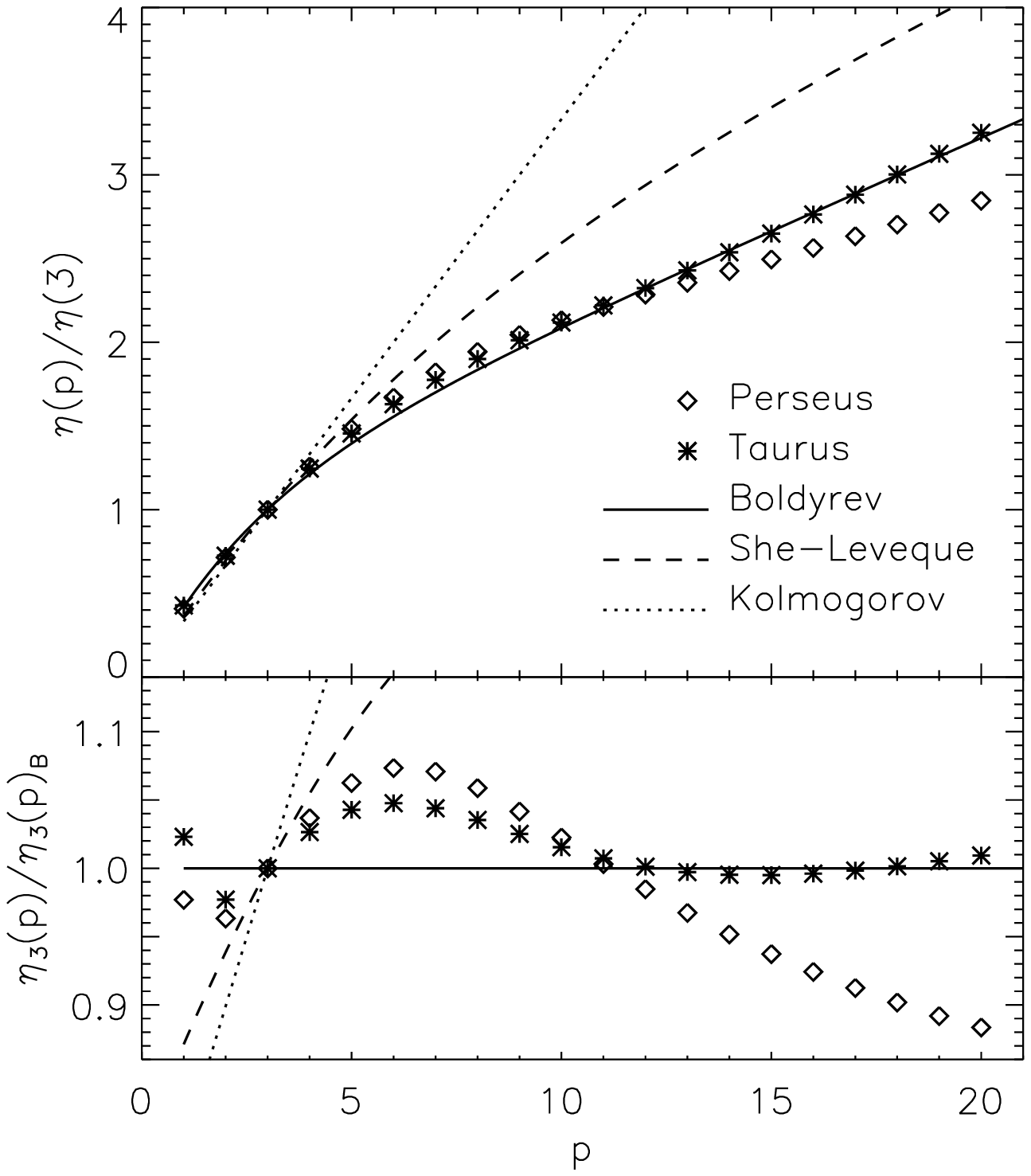}}
\caption[]{}
\label{fig2}
\end{figure}

{\bf Figure \ref{fig2}:} Top panel: Structure function scaling exponents
relative to the third order, $\eta_3(p)=\eta(p)/\eta(3)$, up to the 20th
order, for Taurus (asterisk symbols), Perseus (diamond symbols),
Boldyrev's velocity scaling for super--sonic
turbulence (solid line), She--L\'{e}v\^{e}que's velocity scaling for
incompressible turbulence (dashed line) and Kolmogorov's
velocity scaling (dotted line).
The moments of the structure functions of the integrated intensity images
of Taurus and Perseus follow very closely Boldyrev's
velocity scaling, within 5\% accuracy up to the 20th order (and beyond) for
Taurus and within 10\% accuracy for Perseus. Bottom panel: Same as top
panel, but relative to Boldyrev's scaling,
$\eta_3(p)_{BSL}$. \\

\begin{figure}
\centerline{
\epsfxsize=10cm \epsfbox{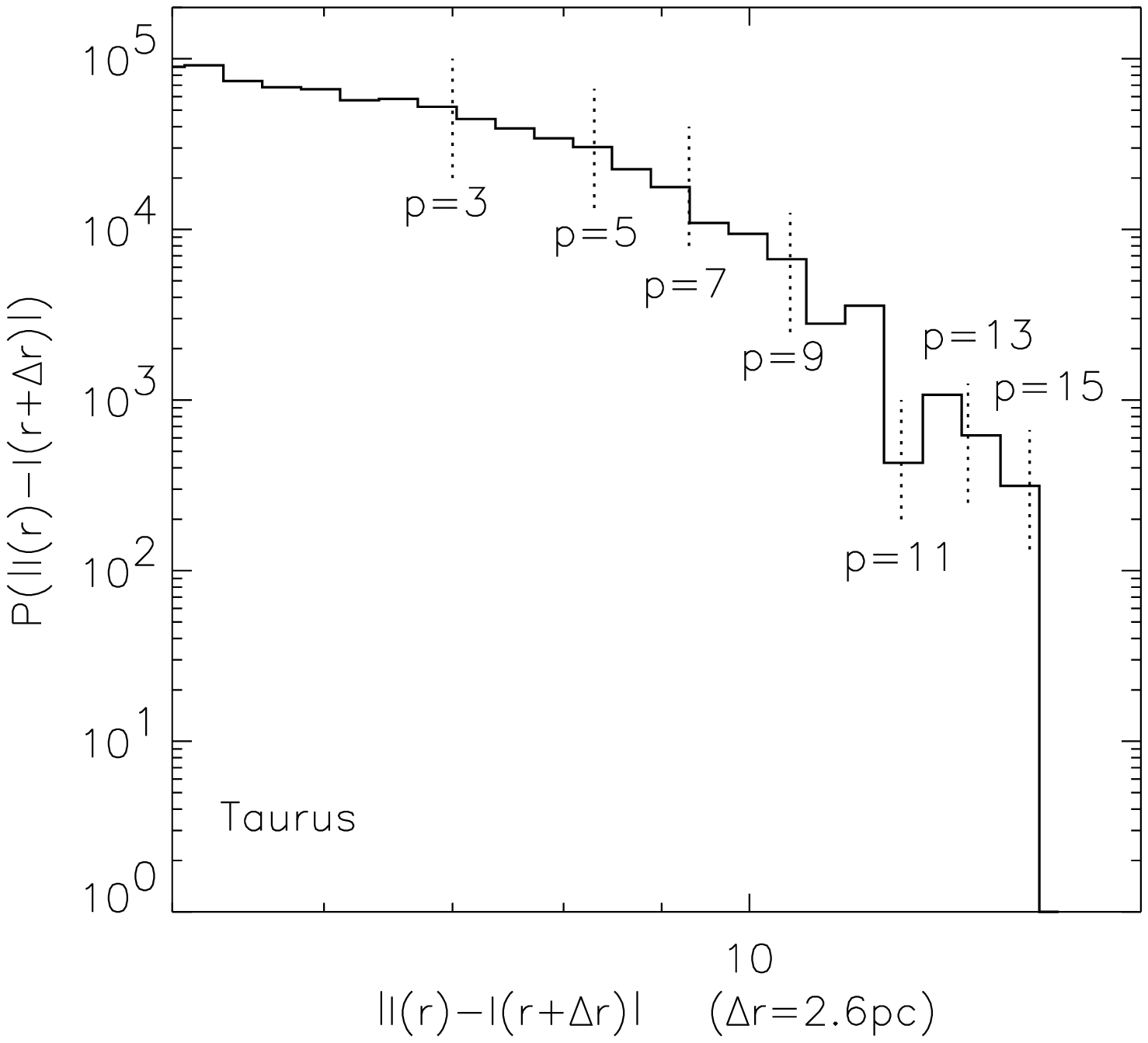}
\epsfxsize=10cm \epsfbox{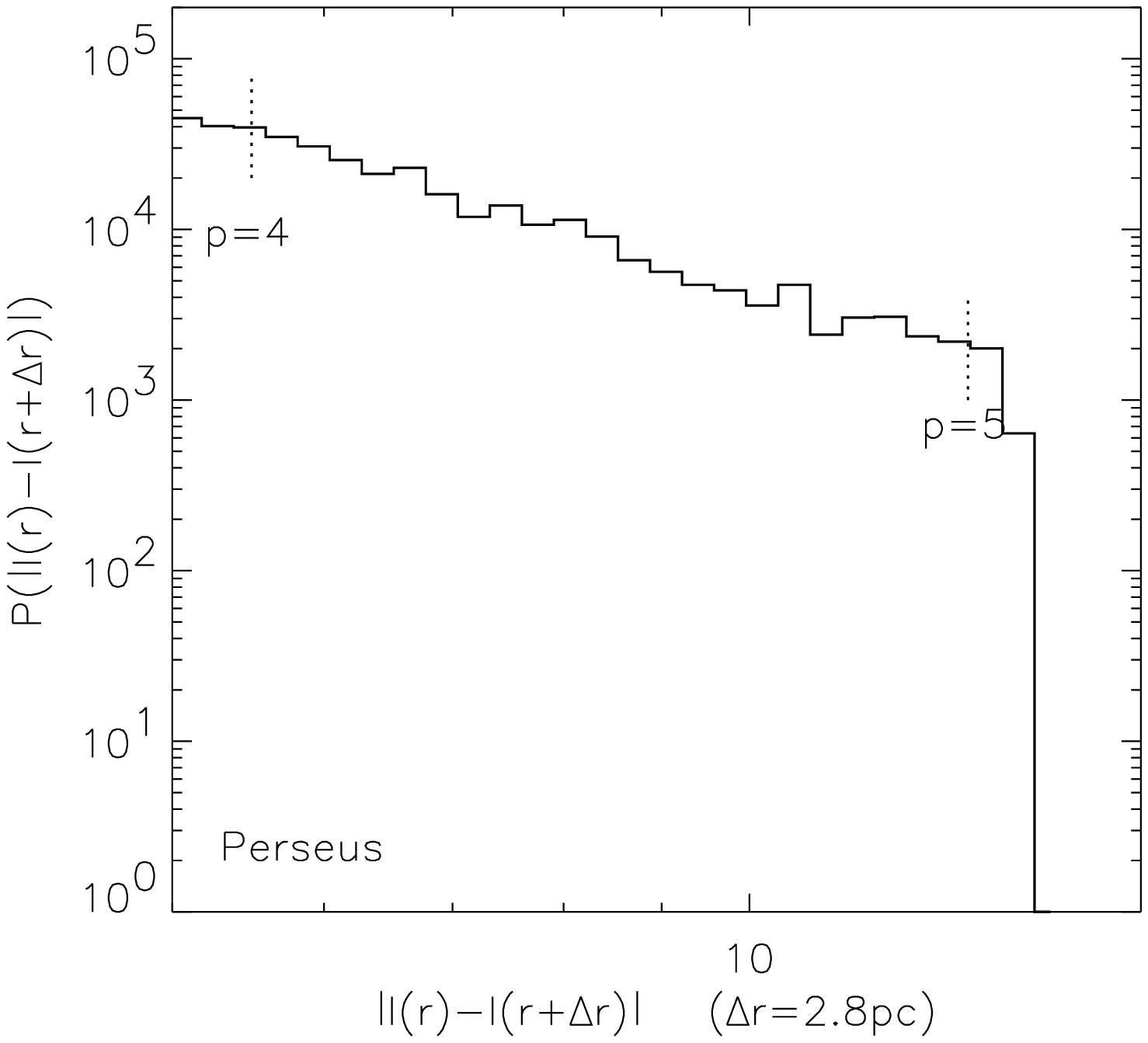}
}
\caption[]{}
\label{fig3}
\end{figure}

{\bf Figure \ref{fig3}:}  Left Panel: Probability density function
(PDF) of the integrated intensity differences of the Taurus
map. Differences are computed at $\Delta r=2.6$~pc. Vertical
dotted segments show the position of the peak of the integrand
of (\ref{pdf}) for orders $p=3$ to 15. The PDF defines well
all moments up to the 15th order. Higher moments are converged due
to the steep cutoff of the PDF. Right panel: Same as left panel,
but for the Perseus map and for $\Delta r=2.8$~pc. The PDF of
Perseus has a very sharp cutoff at large values, which causes
a fast convergence of high order moments. The PDF of Perseus
is such that the peak of the integrand of (\ref{pdf})
is not very pronounced and a large interval of values of the
PDF give therefore a significant contribution to the integral. \\

%{\bf Figure \ref{fig4}:} Top panel: Fluctuations of the structure
%functions as a function of the sample size, relative to the
%structure functions of the largest sample, for orders $p=1$,
%10 and 20 and $\Delta r=0.65$~pc, for the Taurus map.
%The convergence is faster for the structure functions at
%larger values of $\Delta r$. Bottom panel: Same as top panel,
%but for the Pesreus map, and for $\Delta r=0.7$~pc. Also for
%Perseus the convergence is faster for larger values of
%$\Delta r$, since the size of the statistical sample
%grows $\propto \Delta r$.   \\

\begin{figure}
%\centerline{
\epsfxsize=12cm \epsfbox{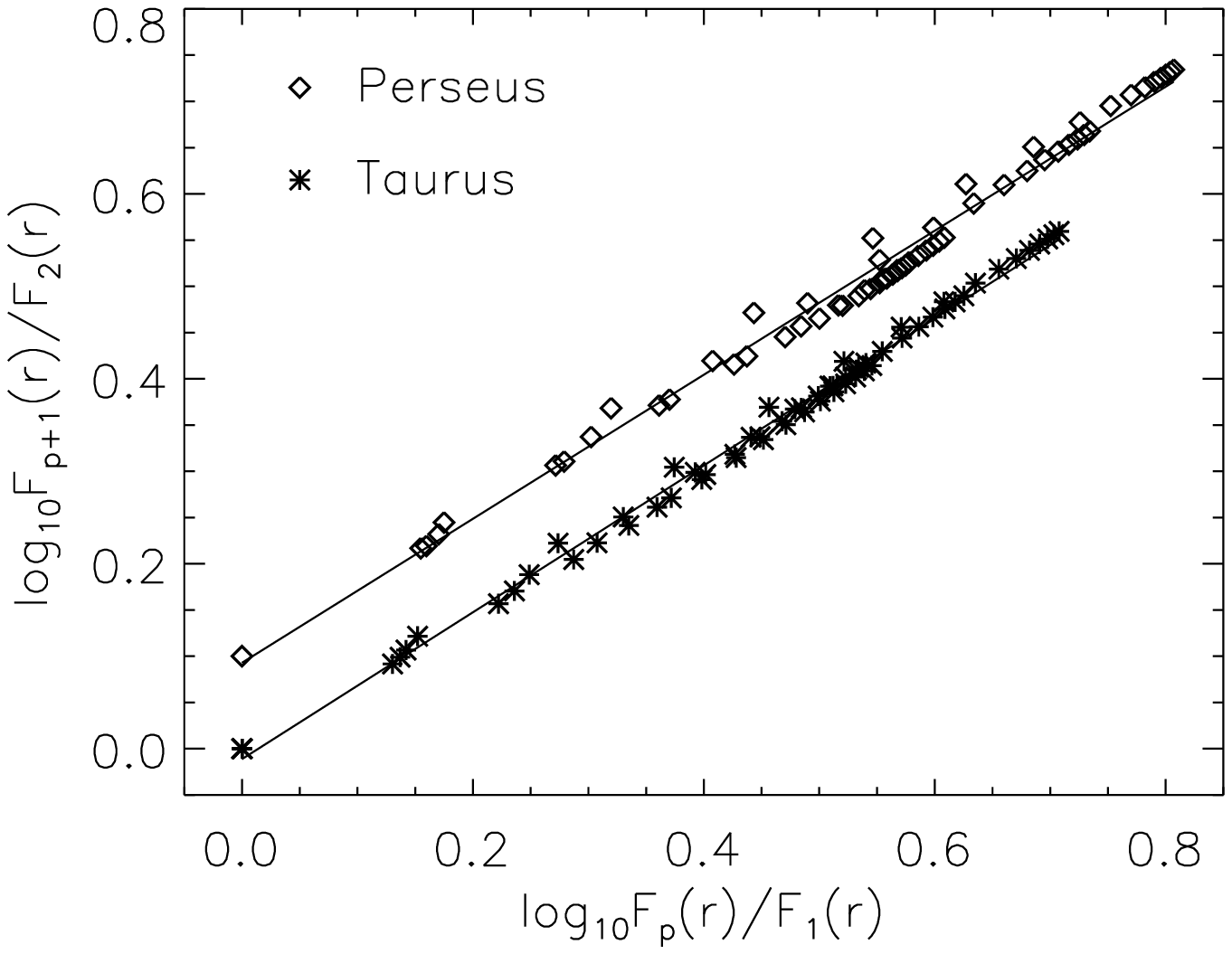}
\epsfxsize=12cm \epsfbox{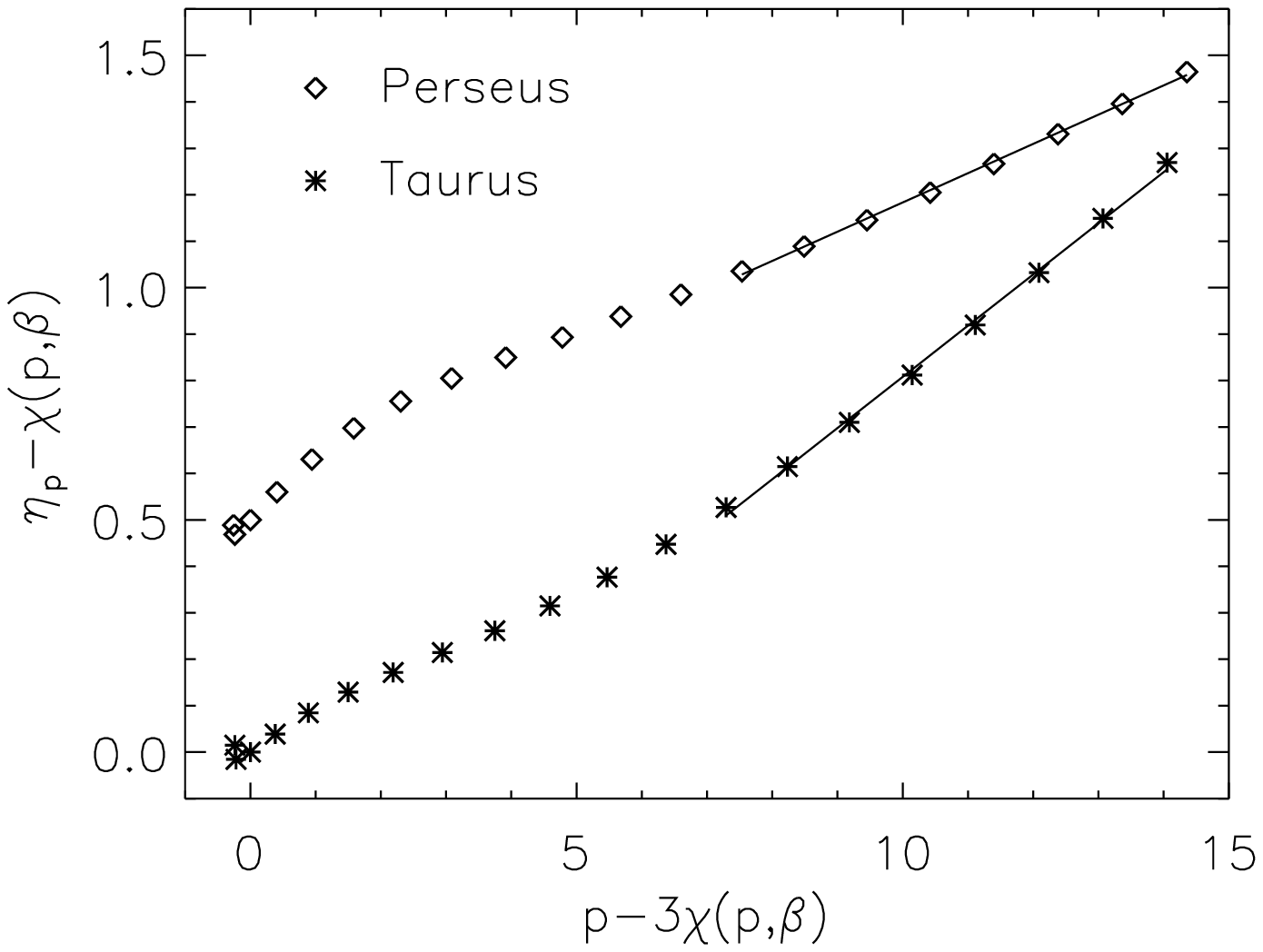}
%}
\caption[]{}
\label{fig5}
\end{figure}

{\bf Figure \ref{fig5}:} Top Panel: The $\beta$ test
(see \S~4.1) that verifies the validity of the hierarchical
structure model. Intensity of fluctuations from both Taurus
and Perseus are well fit by a straight line in this log-log
plot. The data therefore pass the $\beta$ test. The value
of the parameter $\beta$ given by the slope of the least square
fits is $\beta=0.79$ for Taurus and $\beta=0.78$ for Perseus.
Bottom panel: The $\gamma$ test (see \S~4.1). The data is
well fit by a power law at high orders. The solid lines are
least square fits for orders within the range $p=12-20$.
The data therefore pas the $\gamma$ test. The slope of the
least square fits provides the values $\gamma=0.11$ for Taurus
and $gamma=0.06$ for Perseus. \\

\clearpage

\clearpage

\clearpage

\clearpage

\end{document}